# Deep mouse brain two-photon near-infrared fluorescence imaging using a superconducting nanowire single-photon detector array

Amr Tamimi[1], Martin Caldarola[2], Sebastian Hambura[1], Juan C. Boffi[1], Niels Noordzij[2], Johannes W. N. Los[2], Antonio Guardiani[2], Hugo Kooiman[2], Ling Wang[1], Christian Kieser[1], Florian Braun[3], Andreas Fognini[2], Robert Prevedel[1,4-7]

[1] Cell Biology and Biophysics Unit, European Molecular Biology Laboratory, Heidelberg, Germany.
[2] Single Quantum B.V, Delft, The Netherlands.
[3] Chemical Synthesis Core Facility, European Molecular Biology Laboratory, Heidelberg, Germany.
[4] Developmental Biology Unit, European Molecular Biology Laboratory, Heidelberg, Germany.
[5] Epigenetics and Neurobiology Unit, European Molecular Biology Laboratory, Rome, Italy.
[6] German Center for Lung Research (DZL), Heidelberg, Germany.
[7] Interdisciplinary Center of Neurosciences, Heidelberg University, Heidelberg, Germany.

Correspondence: M.C. (m.caldarola@singlequantum.com) or R.P. (prevedel@embl.de).

**Abstract:**

Two-photon microscopy (2PM) has become an important tool in biology to study the structure and function of intact tissues *in-vivo*. However, adult mammalian tissues such as the mouse brain are highly scattering, thereby putting fundamental limits on the achievable imaging depth, which typically resides around 600-800µm. In principle, shifting both the excitation as well as (fluorescence) emission light to the shortwave near-infrared (SWIR, 1000-1700 nm) region promises substantially deeper imaging in 2PM, yet has proven challenging in the past due to the limited availability of detectors and probes in this wavelength region. To overcome these limitations and fully capitalize on the SWIR region, in this work we introduce a novel array of superconducting nanowire single-photon detectors (SNSPDs) and associated custom detection electronics for the use in near-infrared 2PM. The SNSPD array exhibits high efficiency and dynamic range, as well as low dark-count rates over a wide wavelength range. Additionally, the electronics and software permit seamless integration into typical 2PM systems. Together with a fluorescent dye emitting at 1105 nm, we report imaging depth of > 1.1mm in the in-vivo mouse brain, limited only by available labeling density and laser power. Our work further establishes SWIR 2PM approaches and SNSPDs as promising technologies for deep tissue biological imaging.

**INTRODUCTION**

Light microscopy provides a non-invasive and high-resolution, optical way to study biological structure and function. However, in many mammalian tissues light attenuation, i.e. scattering and absorption, poses a grand challenge that often prevents investigations of cells and processes seated inside deep, yet physiologically relevant, tissues. Over the past decades, multi-photon excitation and in particular, two-photon excitation microscopy (2PM), has become the gold standard for recording cellular structure and function inside scattering tissues such as the mouse brain in-vivo [1,2]. However, the maximum penetration depth of two-photon microscopes is fundamentally limited by the onset of out-of-focus fluorescence near the surface with increasing excitation power, and typically reached at 600-800µm depth inside the mammalian brain [3]. Higher-order fluorescence excitation, notably three-photon excitation microscopy (3PM), has shown potential for deeper imaging beyond 1 mm [4–8], yet the significantly lower 3P cross-section compared to 2P demand diligent optimization of excitation laser sources and associated parameters in order to prevent potential photodamage [5,9,10]. An alternative and potentially more promising approach is to further exploit and leverage the red-shifted SWIR wavelength region not only for excitation, but also for fluorescence detection. This in turn necessitates advances in specialized detector technologies, as current InGaS based PMTs have fairly poor quantum efficiency (Q.E. <2.5%) and high dark counts (~$10^5$ $s^{-1}$) compared to visible PMTs.

Recently, high gain, low noise, and high efficiency detectors based on superconducting nanowires single photon detectors (SNSPD) have been developed which show unprecedented performance in the SWIR region [11,12]. These properties make them eminently suitable for many applications including near-infrared microscopy. Previous work has successfully demonstrated their use for confocal bio-imaging applications [13,14], however, the quantum dot fluorescence relied on one-photon excitation by a continuous-wave NIR laser hence not providing the intrinsic optical sectioning and scattering resilience afforded by 2P excitation. Furthermore, only ballistic fluorescence was detected through a pinhole, therefore wasting precious signal. For efficient 2P-excitation in deep tissues, low-repetition rate, fs-pulsed lasers have proven to be optimal [3,10]. With such excitation sources, 2/3PM typically operates in the single laser pulse per pixel regime, which in turn necessitates large area detectors with sufficiently high dynamic range to visualize different levels of fluorescence emanating from each individual voxel. As SNSPDs intrinsically operate in the single-photon detection regime and their sensing area is typically very small (~10's of µm), here we developed a novel SNSPD array composed of 36 (6x6) individual detectors. This layout substantially improves the dynamic range but also effective detector area which improves overall light collection efficiency as non-ballistic fluorescence photons can also be captured.

To fully capitalize on these novel class of SWIR detectors, we synthesized a near-IR organic dye [15] that can be 2P excited at 1700 nm and whose emission at ~1100 nm overlaps with the sensitivity and Q.E. peak of the SNSPD array (**Fig. 1**). Together with the SNSPD array and custom read-out electronics, we demonstrate deep tissue in-vivo microscopy of the adult mouse brain vasculature down to approximately 1.1mm depth, which surpasses traditional 2P depth limits and compares very favorably with more elaborate 3PM experiments in the literature [4–8]. Our work establishes SNSPD arrays as versatile, easy-to-use detectors for 2PM and paves the way for further developments of long-wavelength fluorophores and excitation as well as detector technologies.

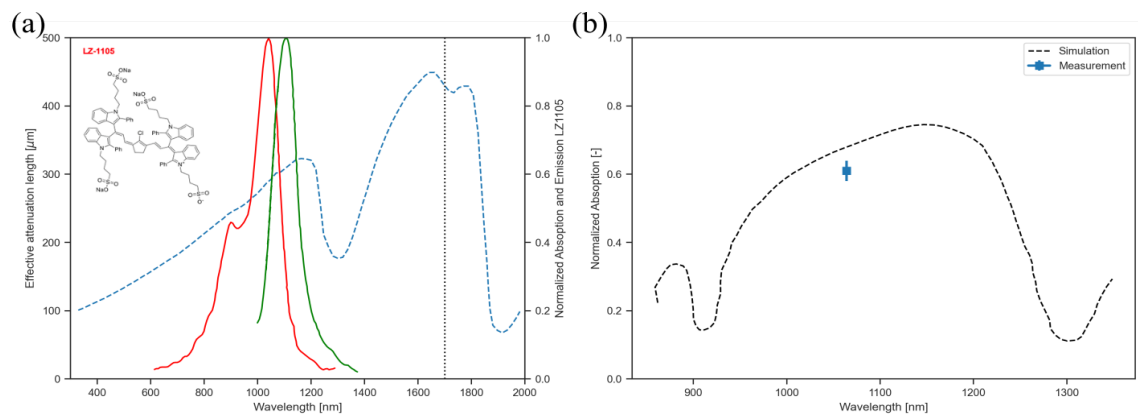

*Figure 1: Spectral characterization of NIR dye and SNSPD array. (a) The blue dashed line depicts the effective attenuation length (left axis) for water, which accounts for the predominant contribution for brain tissue light propagation and highlights the benefits of the SWIR regime. Data extracted from Ref. [16]. The vertical dashed line depicts the laser source used for two-photon excitation while the solid curves depict the (normalized) absorption and emission spectra of the LZ1105 dye (right axis). LZ1105 spectra taken from Ref. [15]. (b) Spectral characterization of the SNSPD array. The dashed curve is a numerical simulation of the expected spectral response of the array of SNSPDs while the square represents the measured array efficiency at 1064 nm. Note that the simulation is usually an upper bound for the efficiency since it does not incorporate material imperfections.*

**RESULTS**

**Development, characterization and integration of the SNSPD array**

Superconducting nanowire single-photon detectors (SNSPDs) were pioneered as a new light detection technology in the SWIR region and have so far found numerous applications in quantum communication and quantum optics [17]. They are based on superconducting nanowires and combine outstanding detection efficiency with very high time resolution and low noise [11], that vastly surpass photomultipliers and APDs in the wavelengths of interest (1000-2500 nm) in terms of sensitivity and time resolution.

By far, the most common implementation of SNSPDs consists of a fiber-coupled system, where the light to be sensed is delivered to the SNSPD inside the cryostat where the superconducting detectors are placed by means of a single-mode optical fiber. For the specific case of 2P microscopy such implementation would be disadvantageous since the single-mode fiber will limit dramatically the number of fluorescence photons to be collected as they contain a high fraction of scattered light. To overcome this, we decided to develop a free-space SNSPD system, in which the light reaches the detectors inside the cryostat through an optical windows Such a system benefits from the high-efficiency and low dark noise of SNSPDs which is instrumental to image faint fluorescence signals deep inside highly scattering tissues such as the mouse brain.

The SNSPD system used in this work consists of a closed-cycle cryostat to cool down the superconducting nanowires to ~3K with the important addition of windows for free-space optical access. Inside the cryostat, we fabricated an array detector which consists of $6 \times 6$ individual SNSPDs with square shape and 10 µm side length, providing a total detection area of 60 µm × 60 µm. The sensitive area of the detector is an important parameter for two-photon excited microscopy as it increases the collection efficiency of the emitted photons, including the ones that experience a high number of scattering events in the tissue and therefore will reach the detector plane in a different position respect to the ballistic photons. Figure 2 shows a picture of the system and a schematic diagram of the SNSPD array and its main elements inside in the cryostat. The bias and readout of each pixel is achieved through independent electrical lines, maintaining the capabilities for single-photon detection and the above-mentioned advantages of SNSPDs in the near-infrared range. However, due to space constraints, only 24 electrical channels can be utilized at present, rendering only 24 out of the 36 pixels of the array active.

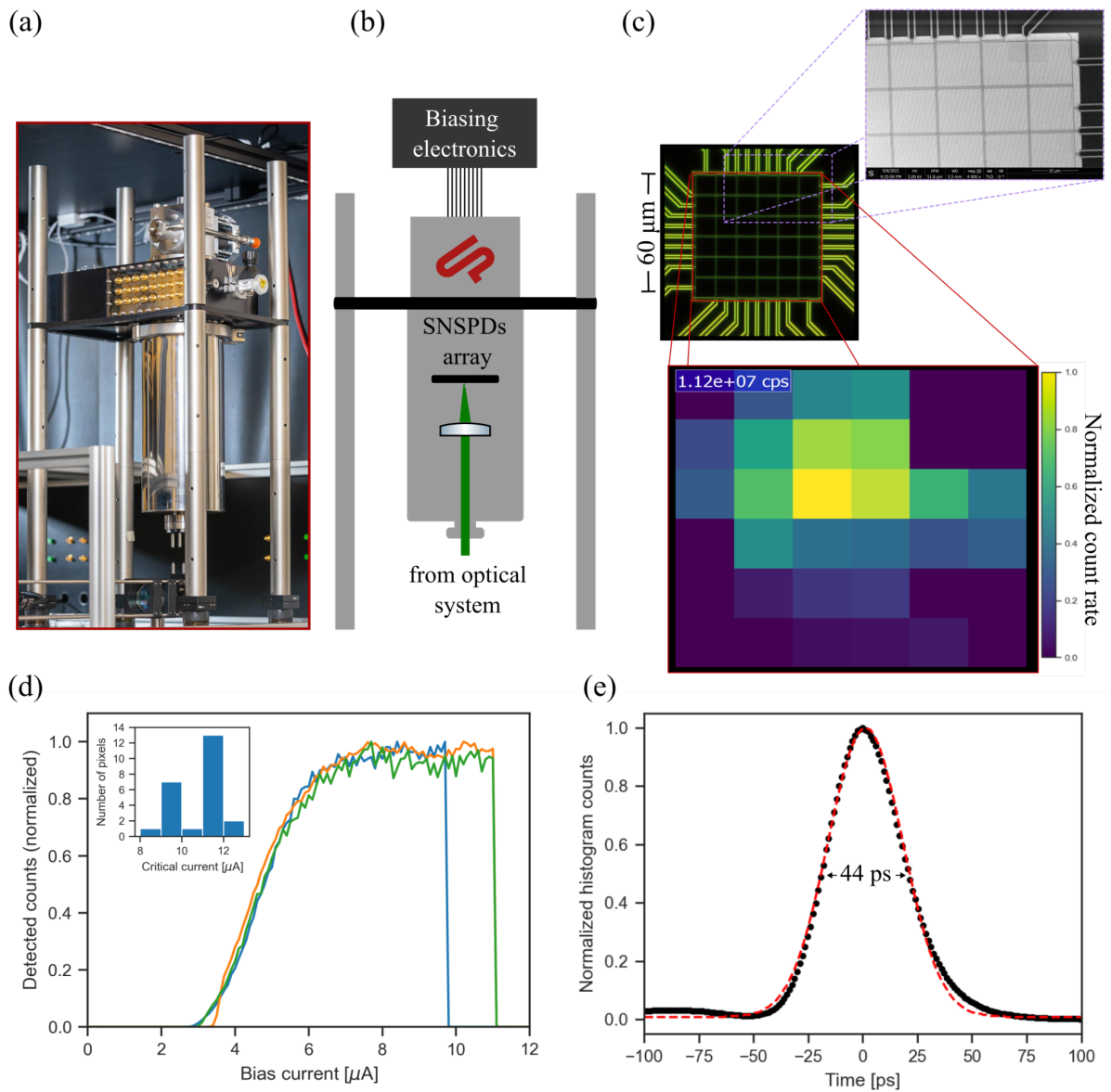

*Figure 2: Schematics and characterization of the free-space coupled SNSPD array.* (a) Optical image of the free-space coupled cryostat. (b) Schematics of the interior of the cryostat, depicting the position of the SNSPD array and the focusing lens and the connection to the control electronics. (c) (center) Optical image (dark field mode) of the 6x6 array with a sensitive area of 60 μm × 60 μm, along with a scanning electron microscope (SEM) image of the top corner of the array (top). (bottom) Map of the detected fluorescence counts using the SNSPD array when incorporated in the 2PE microscope. The colormap shows normalized counts of the Gaussian profile while the inside number depicts the total number of counts detected by the complete array in one second. (d) Bias current sweep of three different pixels from the array, showing a good plateau indicating saturation of the internal efficiency. These curves also provide the critical current at which the SNSPD becomes non-superconducting and the counts go to zero. The inset shows a histogram of the measured critical currents for all the connected elements in the array. (e) Representative timing jitter histogram (black dots), with a FWHM of 44ps (from the Gaussian fit - dashed red line - also see Methods).

A detailed characterization of our SNSPD array system and its main performance metrics is depicted in Figure 2 (c-e). Fig. 2c also depicts typical count rates of the array when 2PE fluorescence from LZ1105 is focused on the center pixel within a size of ~15 μm, which ensures that all the incident light is collected by the complete array. Figure 2 (d) depicts a plot of the normalized detected counts for different bias currents for three representative pixels in the array. The measurements show a clear saturation, indicating a good internal efficiency and reasonable critical currents (tens of μA,) in which superconductivity and thus detection is lost.

We also characterized the SNSPD array quantum efficiency and found a total system efficiency of 57 ± 5% at a wavelength of 1064 nm (see the Methods section for more details). Since the measured transmission for the combined optical windows in the cryostat at 1064 nm is ~0.935 and the transmission of the focusing lens is ~0.99 at 1064 nm, we estimate that the SNSPD array itself has a total detection efficiency of ~61%, as shown in Figure 1 (b). We also characterized the so-called dark-count rate, i.e. the number of detection events when no input light is present, and obtained a total of $9.7 \times 10^3 s^{-1}$ for the complete array, which corresponds to < 300 $s^{-1}$ dark-count rate per pixel on average. Note that a standard photomultiplier tube for the same wavelength range provides a typical quantum efficiency of 2% and comparable dark counts, therefore we expect the SNSPD to yield 2PM images with substantially better signal to noise ratio.

The SNSPD array was then integrated into a custom multiphoton microscope optimized for deep tissue imaging [6,18,19] (see **Fig. 3a** and Methods). In contrast to typical multi-photon microscopy (MPM), however, the fluorescence signal is actively de-scanned by the galvanometric mirrors before being separated from the excitation light using a dichroic mirror and then optically relayed into the cryostat and onto the SNSPD array.

**Electronic interfacing and integration**

In order to have a functional SNSPD, a constant bias current has to flow on the superconducting nanowire. Since the photon-detection event will break the superconductivity temporarily, the bias current is diverted to an amplification stage to generate output pulses of reasonable amplitude for further usage, which is accomplished for 24 individual channels of the 6 x 6 array by a dedicated electronic driver (SQDriver, SingleQuantum). This electronic driver provides an electrical analog output where the amplified pulses coming from the SNSPD can be sent to other electronics, such as a time-tagger or, as in our case, taylor-made electronics aimed to integrate the SNSPDS into existing multi-photon microscopes and their associated hardware control software.

In order to aid the seamless integration of the SNSPD array into common MPMs, we developed custom processing electronics, which provide dual functionality (**Fig. 3b**). First, the raw analog SNSPD driver output pulses, which represent individual photon detection events, are digitized and copied to an FPGA which enables streaming the raw event outputs of all the channels to disk. Additional channels allow various trigger and synchronization signals to be saved simultaneously for offline image reconstruction. This maintains a high temporal (8ns) as well as the full spatial resolution of the fluorescence signal. In parallel, the uniform digital pulses are further signal conditioned with low-pass filters engineered to produce a smooth analogue pulse without ringing (see Methods). The pulses from all 24 channels are summed to represent a single analog signal that can be fed into existing microscopy hardware for direct visualization, effectively replacing the common PMT analog input of a multiphoton microscope. This enables the seamless and unrestricted use of existing MPM data acquisition hardware and related control, visualization and analysis software (NI DAQ and ScanImage2023 [20] in our case).

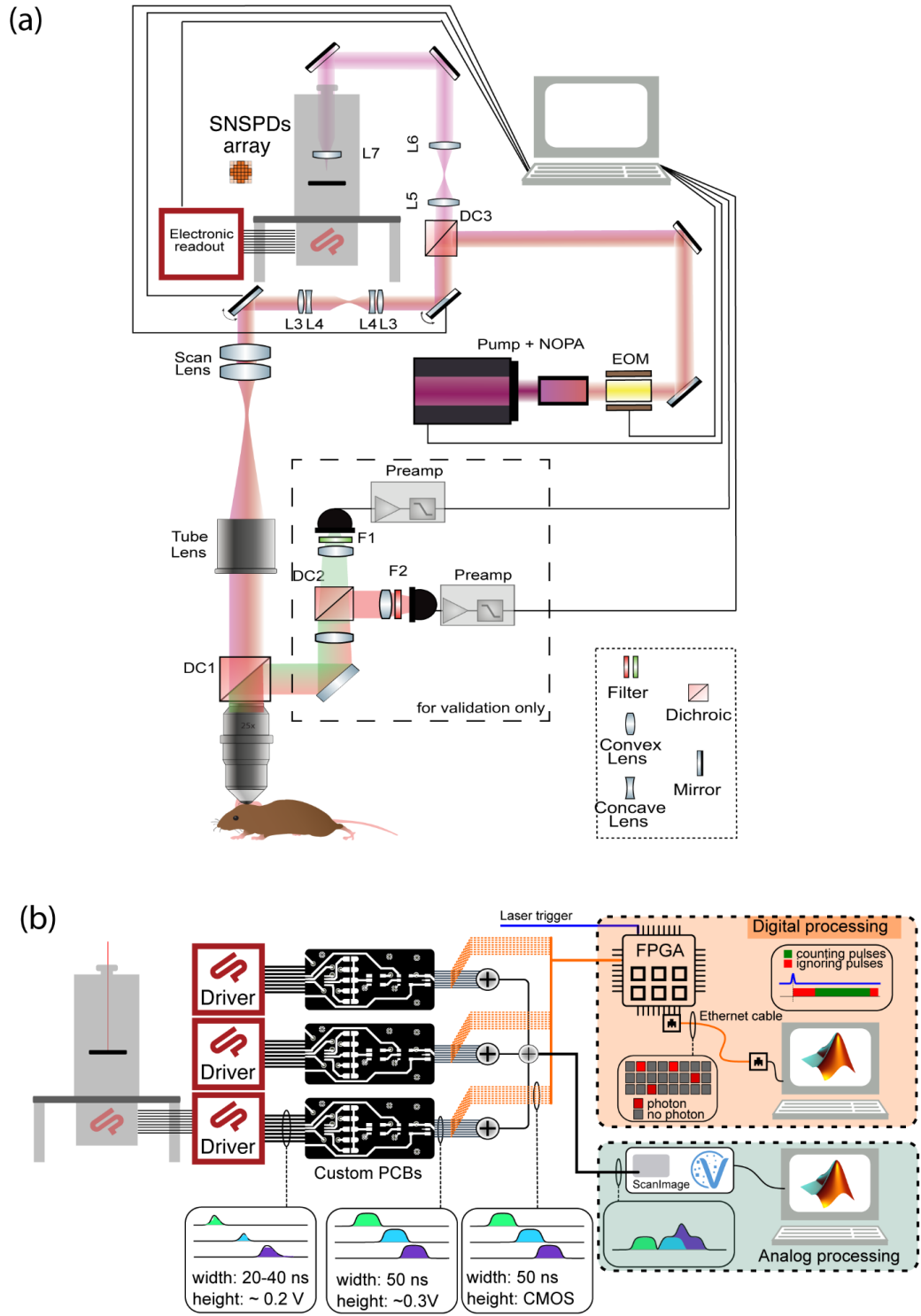

***Figure 3: Integration of the SNSPD array into a custom MPM and dedicated signal processing pipeline.*** *a) Optical layout of the multiphoton microscope. The excited fluorescence is de-scanned and relayed onto the SNSPD array. Note the dashed box containing traditional PMTs was only used during development but is not necessary during imaging. b) Design of the SNSPD signal processing electronics. The 24 outputs of the SNSPD drivers are processed by custom PCBs, and converted into digital signals (3.3V or 5V logic high) as well as analog signals. For the analog processing, the signals are summed into one single analog pulse that existing processing software and hardware can readily use. For the digital processing, a FPGA records the number of pulses (photons) for each channel separately, triggered by the excitation laser, and streamed to the computer for offline image reconstruction (also see Methods for details).*

**Selection of an organic SWIR dye for vascular imaging**

In-vivo imaging in the SWIR region poses the challenge of adequate bio-compatible fluorescent labelling. While several promising nanomaterials do exist [21–23], many of them are either toxic or show poor solubility in water. A common choice is the commercially available dye cardiogreen (ICG), which has demonstrated excellent biocompatibility [24]. However, it only has a relatively short half-life time (~minutes) in the bloodstream, and a diminishing absorption cross-section in the SWIR region above 1000µm [25]. We thus identified the dye LZ-1105 [15] with absorption and emission spectra that overlap well with available SWIR laser sources and our SNSPD array sensitivity (**Fig. 1**). In particular, the absorption spectrum of LZ-1105 shows a prominent shoulder at 900 nm which suggests efficient two-photon excitation at 1700nm. In order to efficiently produce LZ-1105 for in-vivo imaging applications, we developed our own alternative and robust synthesis pipeline for this dye (see Methods). While independent verification of the two-photon cross-section is generally difficult and outside the scope of this work, we found LZ-1105 to be very bright upon 2P excitation with fs-pulsed excitation in vitro and in vivo, with the cross-section not varying significantly between 1650 to 1750 nm.

**In vivo deep brain imaging**

To investigate the suitability of the LZ-1105 dye and explore the performance of the SNSPD array for deep tissue microscopy, we performed in-vivo mouse brain vascular imaging experiments. For this we prepared mice with cranial windows over the visual and motor cortex areas, and performed imaging experiments with the mouse head-fixed and anesthetized (see Methods). An intravenous injection of the synthesized LZ-1105 dye into the mouse tail vein was performed with roughly 100µL of dye solution (in saline buffer) at an appropriate concentration targeting a dye dosage of 5mg/kg. This dosage was considered safe based on Ref. [15], and no adverse effects or evidence of toxicity has been observed.

For our experiments, we utilized 1700 nm pulsed excitation at 1MHz repetition rate and 60 fs pulse duration to efficiently two-photon excite the LZ-1105 dye. Care was taken not to exceed 2nJ per pulse at the focus and 100mW average power to prevent optical damage and/or heating of the brain, respectively [5,10]. With these conditions, we acquired image stacks as shown in Fig. 4, where arteries and veins throughout the cortex can clearly be resolved and the images retain high signal-to-background ratios (SBRs) down to a maximum depth of ~1150µm, i.e. below the cortex, where SBR reaches ~1. This represents the depth limit for our SNSPD array based MPM. At the deepest imaging depth of ~1.1mm, the laser power was ~100mW with an integration time of approximately 10 sec. We note that the achievable imaging depth was predominantly limited by the available vascular labeling density and laser power, and that in principle larger imaging depth could be obtained for sparser and/or higher concentration labelling.

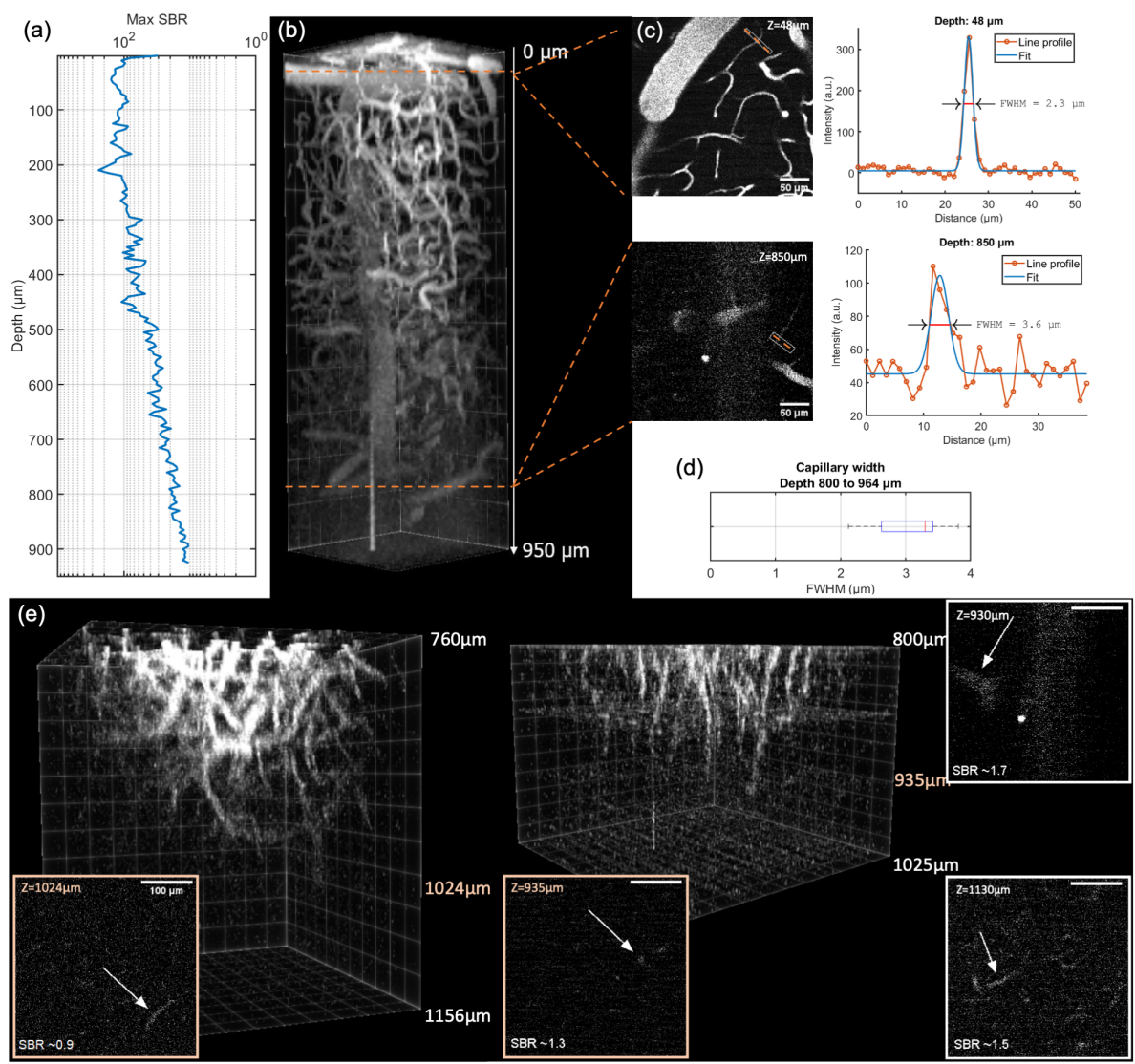

*Figure 4: Experimental in-vivo deep brain imaging with a custom SNSPD array multiphoton microscope. (a-b) 3D image-stack of an adult mouse brain vasculature down to a maximum depth of 950 µm (from the brain surface), and corresponding signal-to-background ratio (SBR) throughout the image volume. The stack consists of 186 slices at 5µm z-intervals. Each slice is 256x256 pixels and covers a ~300µm field of view. Dwell time is 6.55 seconds per z slice (10 frames averaged, 10 laser pulses per pixel). (c) Exemplary slices at 48 µm and 850 µm depth showing high SBR and capillary details. Line profiles over blood vessels are shown on the right, with a fit that indicates achievable spatial resolution. (d) Statistical quantification of capillary vessel diameter at large image depth (800-964µm; n=6), indicating an upper bound of the achievable lateral spatial resolution at this depth. (e) Deep image stacks covering vasculature between 760 and 1156µm, indicating the highest possible imaging depth at which SBR falls to ~1. Arrows point to vascular structures. Exemplary data from four different experiments using three different mice.*

**Experimental investigation of depth limit and spatial resolution**

To verify efficient collection of fluorescence throughout the entire imaging depth, we also analyzed the photon count distribution on the SNSPD array. We found that the center pixel of the array on average receives around 8000 counts (photons) per second, which is well within the previously characterized optimal linear response regime of 10-50k counts. For single photon counting with a pulsed source at 1MHz repetition rate, and given that this is an average rate over the entire image frame, we estimate that the brightest spots in the sample generate a count rate of ~80-100k photons per second. In the case of multi-photon excitation, in particular, relatively lower repetition rate lasers are necessary (0.5-

10MHz). For efficient detection, it is thus essential to spread the signal over multiple pixels and the array detector provides the dynamic range that makes this kind of deep imaging experiment possible.

To characterize the achievable spatial resolution, we analyzed the narrowest blood vessels in each image frame, establishing an upper bound estimation for the lateral resolution (**Fig. 4c**). Specifically, two slices were selected, one at a superficial depth of 60µm and another at 850µm depth and line profiles are plotted across the vessel. We also note that this procedure itself is conservative since blood vessels are in general larger (4.2 ± 0.4µm, [26]) than the lateral extent of the excitation PSF. On average, we found an average vessel width of 3.0 ± 0.6µm at an average depth of 856µm (**Fig. 4d**). Therefore, we conclude that our SNSPD detection system does not negatively affect the achievable spatial resolution and that our combination of low-noise sensitive detectors with multi-photon excitation enables high resolution imaging at advanced depths in the infrared fluorescence range.

**Gated digital counting improves image SBR**

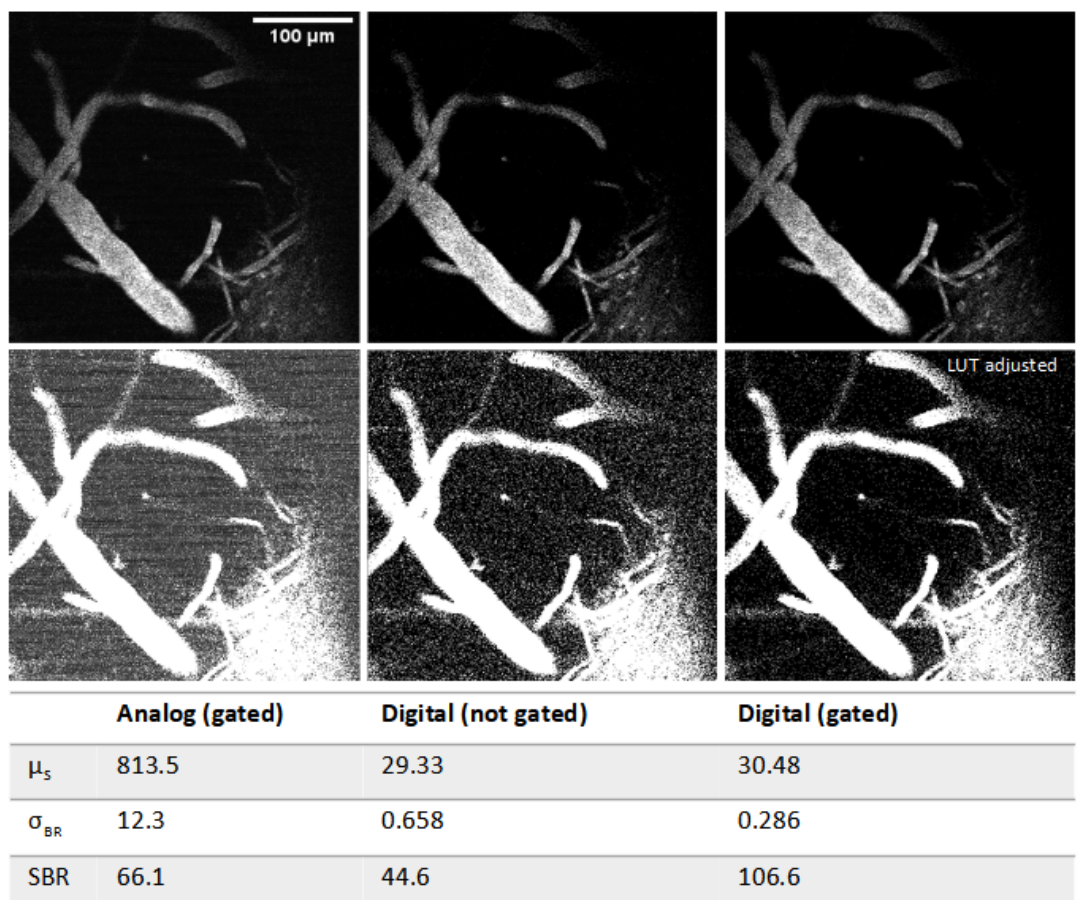

|  | Analog (gated) | Digital (not gated) | Digital (gated) |
|---|---|---|---|
| $\mu_S$ | 813.5 | 29.33 | 30.48 |
| $\sigma_{BR}$ | 12.3 | 0.658 | 0.286 |
| SBR | 66.1 | 44.6 | 106.6 |

*Figure 5:* The effect of analog and digital gating on image contrast. The same image, showing clear signal and background regions, is acquired to compare the achievable SBR between three different modalities. The top row shows the resulting image in which (left) analog summation, (middle) digital counting without and (right) with gating (80ns width). The bottom row has the contrast adjusted for better visualization of the noise floor. The best SBR is achieved in gated digital mode. $\mu_S$, mean of signal, $\sigma_{BR}$, standard deviation of background.

The custom electronics developed for the SNSPD array allow not only an analog summation of the total signal detected by the 24-pixel channels, but to also digitize them to be used by a FPGA module which provides counting functionality via a user-defined delay and integration period (see Methods). This allows reconstruction of the image from the binary counting information, and adds additional capability to further improve the image SBR by suppressing background noise. We characterized the effects of analog vs. counting (digital) mode as well as the role of gating in suppressing background noise. For this an image at 42µm depth was acquired with three different modalities, i.e. with analog signal summation, digital counting with and without gating (**Fig. 5**). In the digital mode without gating, all events (photons) are recorded between excitation pulses, leading to the integration of both fluorescence and dark counts or stray light (background noise). The gated image in turn was acquired in counting mode with the gate set 80ns width following each excitation pulse. As evident from Fig. 5, gated digital counting improves the SBR by ~60%. This improvement is mainly due to the suppression of background analogue ripple that is caused by RF noise and ground loops, but also due to other analogue noise sources (e.g. ADC quantization noise). There is an even greater improvement observed in going from ungated digital to gated digital (~139%), which is attributed to the effective suppression of dark counts and stray light. Therefore, digital counting mode achieves the best possible SBR at depth.

**DISCUSSION AND CONCLUSION**

To summarize, in this work we developed and characterized a custom multi-element SNSPD array for bio-imaging applications. To the best of our knowledge, this is the first time that a dedicated SNSPD array technology has been developed and integrated into a multi-photon microscope. Together with the synthesis of a near-infrared dye emitting at ~1100nm, this enables deep-tissue microscopy via two-photon excited fluorescence in the SWIR region. In this context, the multi-element SNSPD array provides the following distinct enabling features: It provides increased dynamic range as required by a low-repetition rate, pulsed excitation regime as typically necessary in deep-tissue microscopy. Here, compared to standard PMT detection, the SNSPD array provides the low-noise and dark counts necessary for high SBR in low (fluorescence) signal settings, which can further be enhanced by digital gating and postprocessing as enabled by custom electronics. It is noteworthy that here we achieve tissue imaging depth over 1mm that previously could only be realized by more advanced and elaborate three-photon excitation modalities [4–6]. Although not demonstrated here, we further note that the SNSPD array and its electronics developed in this work could readily be utilized in other challenging bio-imaging applications such fluorescence lifetime microscopy. Furthermore, the fact that the array records the spatial position of the detected photon could be further exploited for image scanning microscopy [27,28] and related approaches that yield improved spatial resolution, or for other methods aimed to improve image contrast by selectively rejecting background or out-of-focus photons in post processing. Altogether, our work paves the way for more efficient near-IR multi-photon microscopy and may motivate further work on high-gain, and low-noise SWIR detectors as well as near-IR fluorescent probes tailored for biological imaging.

**MATERIAL AND METHODS**

1. **Custom MPM setup**

The core hardware of the 2P near-IR microscope has been described in detail in previous work [6]. In the following, a brief summary of the instrumentation is provided emphasizing any differences from the details given in Ref. [6]: As a laser source, we also employed a Class 5 Photonics White Dwarf WD-1300-dual laser. The 1700 nm channel used here provided a maximum of over 7 Watts at a repetition rate of 1 MHz. A motorized half-wave plate followed by a polarizer, in addition to a reflective optical density filter with a static OD=0.8 attenuation was used to adapt the power range of the laser, yielding a maximum of 100 mW after the objective. Dispersion pre-compensation was done by an internal module in the White Dwarf which yielded 60 fs pulses after the objective (Olympus, 25x NA1.05 water immersion). The custom MPM is controlled via ScanImage (VidrioTechnologies) and its associated National Instruments data acquisition system.

The SNSPD array was optically coupled to the microscope by directing the fluorescence into the outermost window of the cryostat (see Figure 2 (a)). This is achieved by a long pass dichroic at 1200 nm (Edmund Optics) situated just before the galvanometric mirror scanning system, therefore the fluorescence is effectively de-scanned before it is directed to the SNSPD array. The optical relay to the array is composed of a cage system, protected silver mirrors and a 1:1 telescope (f=150mm; Thorlabs achromats, B-Coated)). Alignment was optimized by visualizing the size of the detection PSF directly on the array, adjusting it to be centered and that the outer detector elements have approximately a factor of ten less counts than the center pixel. This ensures good collection efficiency even in the case of more severe tissue scattering which results in non-ballistic fluorescence and a larger detection footprint on the detector.

2. **Fabrication of the SNSPD array.**

For the deposition of the superconducting NbTiN film an AJA sputtering machine was used, configured to co-sputter Nb and Ti. Contacts, routing, and alignment markers were defined in Au using a laser writer and using a standard lift-off process. Subsequently the meander pattern was patterned using an EBPG and the positive tone AR-P 6200 resist. After patterning the NbTiN was etched using an SF6 and oxygen plasma. It is important to note these samples are sensitive to ESD discharge, so long SEM imaging should be avoided, and during handling measures should be taken to prevent ESD discharge.

3. **Characterization of the SNSPD array**

A full characterization of an SNSPD consists of the measurement of the quantum efficiency, which states the probability of detecting an incident photon, the dark count rate, the timing resolution (jitter).

To measure the quantum efficiency of the system, a known number of photons in a given time interval is sent to the system with a linear polarization parallel to the SNSPD meander, which maximizes the detected counts. By comparing the incident photon rate with the detected photon rate experimentally determine the detection efficiency or the quantum efficiency of the system. Note that for this procedure we treat the detector as a bucket detector, i.e. we sum the detected counts for all the pixels while making sure that the incident photons are all arriving on the sensitive area of the array. For determining the dark count rate, we turn off all light sources and cover the SNSPD system and measure the number of counts detected.

We also characterized the time resolution for single-photon detection using the standard procedure for jitter determination of an SNSPD. Briefly, we use a pulsed laser to synchronize our laser and detection and record a histogram of detected times. The width of this histogram is a measure of the timing jitter for the detector (see Fig. 2e).

4. **Electronic interface**

The SNSPD drivers generate a ~20 ns long and about 200mV peak pulse for each detected photon. In order to enable high-quality data processing and to simplify the downstream analysis we implemented the following pulse processing: The output signal of the driver is cleaned from high-frequency noise by a first low-pass filter. Then, custom-made PCBs standardize the voltage level using a comparator with a user-adjustable threshold, and the pulse duration using monoflops, ensuring fixed-length pulses independent of the incoming pulse length. This produces reliable pulses with a 50 ns width, and a height of either 3.3V or 5V (user-selectable) for every photon that hits the SNSPD-array. These signals can be interpreted as digital pulses, and are fed directly to the FPGA (Cora Z7-07S) for digital processing. For the analog processing, the 24 channels are scaled down to roughly 300mV, low-passed and then summed together to have a single analog channel going into the existing DAQ system of the MPM (NI-5734 : 4 analog inputs, 80 MS/s 16-bit, absolute maximum voltage ±10 V DC). More details are provided in SI Fig. 1.

The electronics discussed above and in Fig. 3b not only yield an analogue summation of the total signal detected by the 24-pixel channels of the current SNSPD array, but also make all those individual channels available to an FPGA module which provides counting functionality. For every laser pulse of the excitation laser, the FPGA module resets an on-board counter which forms the detection time-base. This counter is incremented at the FPGA clock rate of 125MHz to yield time bins of 8ns each. A user-defined delay and integration period then form the parameters for an integration gate during which the

FPGA will have the channels armed to record a photon arrival. Within the gate, up to one photon can be recorded (an extremely long digital pulse can be optionally recorded as two photons), and outside the gate nothing is recorded. This yields an array of logical 1 and 0 associated with each pixel and each laser pulse, indicating the arrival of a photon during the integration period associated with that pulse. Some auxiliary pulses (ScanImage's frame and line triggers) are also recorded by the FPGA. This data is streamed via TCP/IP server to a binary file along with an external frame clock signal, allowing reconstruction of the image from the binary counting information. A rough image reconstruction is done using the auxiliary synchronization lines, and then by cross-correlating each digital image line with the corresponding analog image line the remaining synchronization and pixel alignment is achieved.

5. **Dye synthesis and characterization**

The synthesis of LZ-1105 was achieved in three steps starting from 2-phenyl-1H-indole using a modified protocol of literature known procedure [15] to obtain 4-(2-phenyl-1H-indol-1-yl)butane-1-sulfonate FB25-06 in the first step after alkylation with 1,2-oxathiane 2,2-dioxide. The key for the synthesis of LZ-1105 was the stepwise reaction of FB25-06 with acetyl chloride in the presence of acetic anhydride in toluene to afford the dimeric compound FB25-21 in very good yield. The following reaction with readily available N-((E)-(2-chloro-3-((E)-(phenylimino)methyl)cyclopent-2-en-1-ylidene)methyl)anilinium chloride (FB25-24) [29] in methanol in the presence of hydrochloric acid provided the target dye.

Reagents were purchased from Sigma Aldrich (Germany) and TCI (Belgium) and used without further purification. All solvents, including anhydrous solvents, were used as obtained from the commercial sources. Air and water-sensitive reagents and reactions were generally handled under argon atmosphere. The reaction progress was monitored by TLC on Merck silica gel plates 60 F254. Detection was executed with a UV-Kabinett HP-UVIS (biostep) at 254 nm or with potassium permanganate staining. Flash chromatographic purification was performed on a Biotage® Isolera One purification system using Biotage® SFär C18 D flash cartridges. Nuclear magnetic resonance spectra were recorded on a Bruker Avance (400 MHz) NMR System at 298 K. Chemical shifts (δ) are given in parts per million (ppm), coupling constants (J) given in Hertz (Hz) and multiplicity is reported using standard abbreviations. UHPLC/MS analyses were performed on Agilent 1290 series equipment consisting of an Agilent 1290 quaternary pump, a 1290 sampler, a 1290 thermostated column compartment and a 1290 Diode array detector VL+ equipped with a quadrupole LC/MS 6120 and an Infinity 1260 ELSD. The analytical column used was a Titan C18 UHPLC Column (2.1 X 30 mm, 1.9 μm) operated at 40 °C and 1.5 ml/min flow rate with a gradient (10% to 15% B in 0.4 min, 15% to 100% B in 1.6 min, 100% B for 0.5 min) using water (A) and acetonitrile (B), both containing 0.1% TFA, as solvents. Compound purity was determined by ELSD monitoring. N-((E)-(2-chloro-3-((E)-(phenylimino)methyl)cyclopent-2-en-1-ylidene)methyl)anilinium chloride (FB25-24) was synthesized following the literature procedure [29].

6. **Animal preparations and imaging procedure**

This work followed the European Communities Council Directive (2010/63/EU) to minimize animal pain and discomfort. All procedures described in this paper were approved by EMBL's committee for animal welfare and institutional animal care and use, under protocol numbers RP170001 and 22-004_HD_RP. Experiments were performed on male and female, 7–24-week-old C57Bl6/j or homozygous Thy1–EGFP-M (Jax no. 007788) transgenic mice from the EMBL Heidelberg core colonies. During the course of the study, mice were housed in groups of 1–5 in makrolon type 2L or 3H cages, in ventilated racks at room temperature and 50% humidity while kept in a 12/12 h light/dark cycle. Food and water were available ad libitum.

The cranial window implantation surgical procedure has been extensively described elsewhere (cite Lina's paper). Briefly, 7-8 week old mice were anesthetized by i.p. Injection of a mixture of 40 μl fentanyl (0.1 mg/ml; Janssen), 160 μl midazolam (5 mg/ml; Hameln) and 60 μl medetomidin (1 mg/ml; Pfizer), dosed in 5 μl/g body weight. Hair over the scalp was removed with hair removal cream and eye ointment was applied (Bepanthen, Bayer). Anesthetized mice were then subcutaneously injected with 1% xylocain (AstraZeneca) under the scalp as preincisional local anesthesia and placed in a stereotaxic frame (RWD life science, model 68803). The skin and periosteum over the dorsal cranium were removed with fine forceps and scissors to expose the bone. A 4-mm diameter circular craniectomy was made centered over the right visual cortex using a dental drill (Microtorque, Harvard Apparatus, 2.5 mm posterior and 2.5 mm lateral to Bregma). Close care was taken to preserve the integrity of the dura and avoid bleeding. A round 4 mm coverslip (around 170 μm thick, disinfected with 70% ethanol) was placed over the craniectomy with a drop of saline between the glass and the dura. The craniectomy and cranial window were sealed using acrylic dental cement (Hager Werken Cyano Fast and Paladur acrylic powder), and a customized metal headbar was cemented to the skull for head fixation under the microscope. The skin wound around the surgical area was also closed with dental acrylic. After surgery, mice received pain relief (Metacam, Boehringer Ingelheim, subcutaneous injection, 0.1 mg ml$^{-1}$, dosed 10 μl g$^{-1}$ body weight) and anesthesia was antagonized by subcutaneous injection of a mixture of 120 μl sterile saline, 800 μl flumazenil (0.1 mg/ml; Fresenius Kabi), and 60 μl atipamezole (5 mg/ml; Pfizer) dosed in 10 μl/g body weight. Mice were single housed after cranial window implantation and had a recovery period of at least 3 weeks before further experiments to resolve the inflammation associated with this surgery [30].

For imaging, a chronic window implanted Thy1-EGFP-M or WT mouse was head-fixed under the microscope and anaesthetized with isoflurane vapor mixed with O2 (5% for induction and 1–1.5% for maintenance). An intravenous injection into the mouse tail vein was performed with roughly 100μL of dye solution (in saline buffer) at an appropriate concentration targeting a dye dosage of 5mg/kg. This dosage was considered safe based on Ref. [15], and we did not observe evidence of toxicity even after three separate injections at this dosage. Care was taken not to exceed 2nJ per pulse at the focus and average power below 100mW to prevent photo damage to the brain. With these conditions, the imaging stacks in Fig. 4 were acquired.

For image analysis, Fiji [31] was used. For the SBR calculation, blood vessels were segmented using the software package [32]. One frame is minimally hand annotated to train the model to recognize blood vessels, resulting in probability maps for both signal and background for the entire stack. These probability maps are thresholded conservatively (high probability) to generate binary signal and background masks. These masks are then fed into the FIJI ROI editor as ROIs, and measurements defined on them, in the case of the signal, mean and maximum, and in the case of the background, mean and standard deviation. Mean and max SBR figures are then generated by subtracting the background mean from the signal mean and max, respectively, and divided by the background standard deviation. For spatial resolution analysis, 5- or 10-pixel wide line profiles were plotted across blood vessels. The width of the line determines how many pixels are averaged for each point of the profile. We note that this averaging procedure can only result in a wider profile, and that a blood vessel cannot be observed to be narrower than the excitation PSF. Therefore, the calculated resolution is indeed a conservative, upper bound on the PSF size.

**ACKNOWLEDGMENTS**

We would like to thank the mechanical and electronic workshops at EMBL Heidelberg for help, and especially the LAR facility and its staff for animal husbandry and help with in-vivo dye injections. The majority of this work was supported by the European Commission (grant no. 951991, Brainiaqs). Furthermore, R.P. acknowledges support of an ERC Consolidator Grant (no. 864027, Brillouin4Life), the Deutsche Forschungsgemeinschaft (project no. 425902099) and the Chan Zuckerberg Initiative

(Deep Tissue Imaging grant no. 2020-225346). This work was supported by the European Molecular Biology Laboratory.

**AUTHOR CONTRIBUTIONS**

A.T. integrated SNSPD array, acquired and analyzed data. S.H. developed the digital processing pathway. C.K developed the analog processing electronics with inputs from A.F J.C.B. performed animal work and surgeries. F.B. synthesized the LZ-1105 dye. L.W. contributed with optomechanical simulations and design. M.C. designed and built the SNSPD characterization setups. A.G and H.K measured the characterization of the SNSPD array. N.N and N.L designed and fabricated the SNSPD array. A.F. designed part of the electronics interface and supervised the work at Single Quantum. R.P. conceived and supervised the project, and wrote the paper with input from M.C. and S.H.

**COMPETING FINANCIAL INTERESTS**

The authors declare the following competing financial interest(s): The following authors were employed by Single Quantum B.V. and may profit financially: N.N, J.N.L.L, A. G., H.K., M.C, A.F. The other authors declare no competing financial interests.

**Supplementary information**

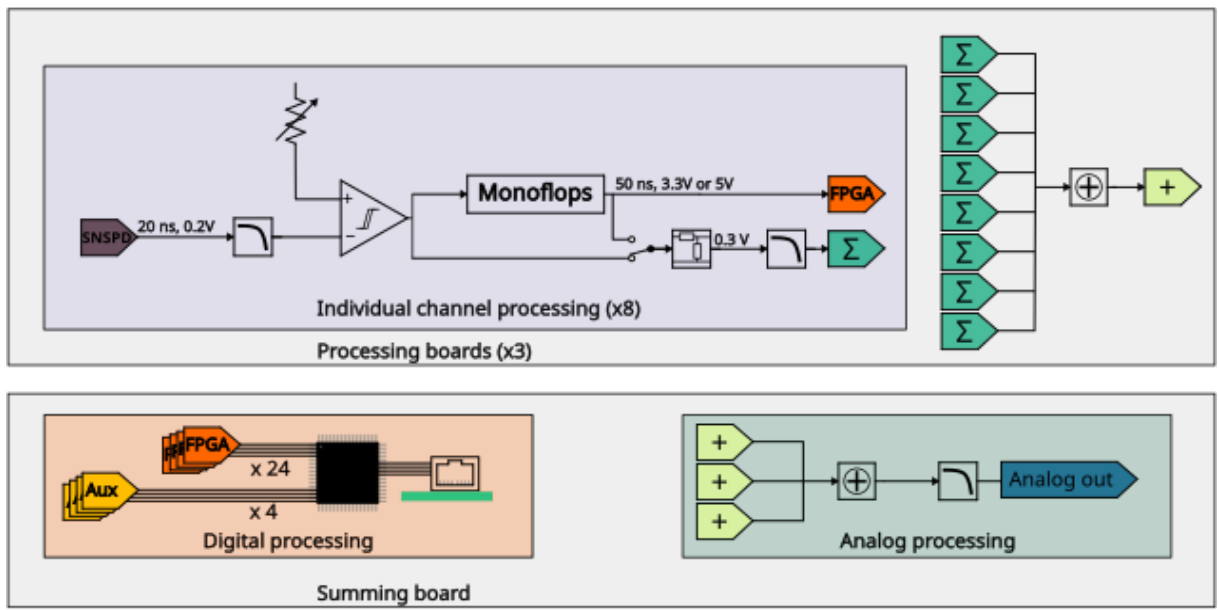

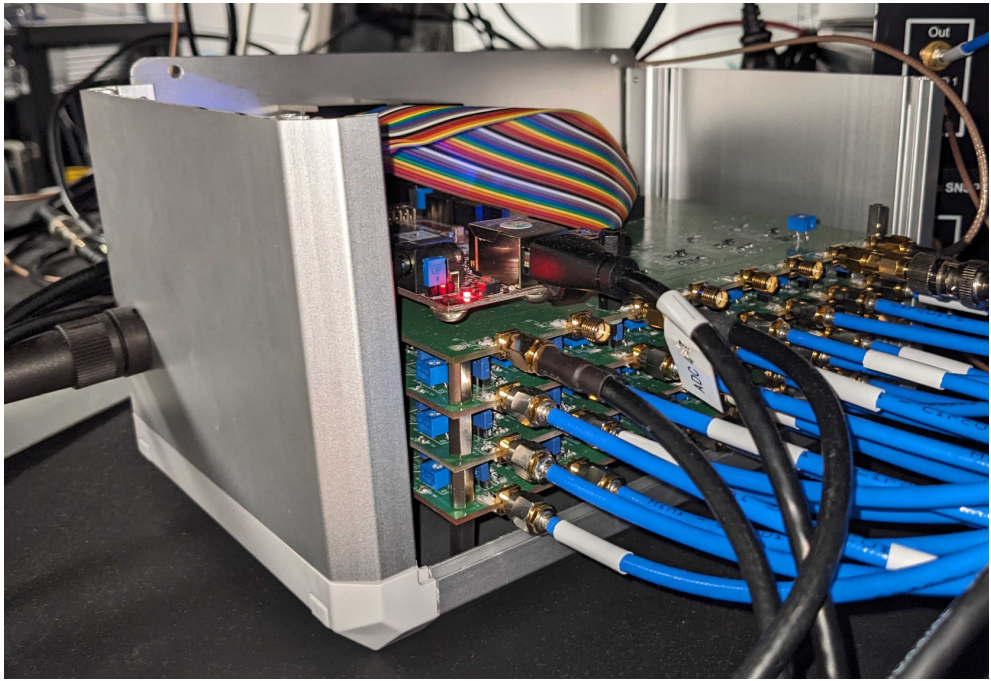

**Figure SI 1: Custom electronic interface.** (Top) The 8-channel signal processing board. 20ns and 200mV pulses are converted into one 50 ns and 300mV pulse (for analog processing) and one 50ns digital pulse (3.3V or 5V logic high) and the 8 analog signals are summed. Three separate boards are combined to allow recording of 24 independent channels. (Middle) A dedicated PCB board sums the analog signal of the 3 previous boards into a single analog output. It also serves as an intermediate between the digital outputs and the FPGA board, allowing only one connector to be used between the FPGA and this setup. The auxiliary signals such as laser triggers are also sampled here. (Bottom) A picture of the physical device where the bottom 3 boards are the processing boards, the summing board directly above, and the FPGA on top.